\documentclass[%
 reprint,
 amsmath,amssymb,
 aps,
]{revtex4-1}

\usepackage{soul}
\usepackage{url}
\usepackage{graphicx}
\usepackage{dcolumn}
\usepackage{bm}
\usepackage{amsmath}
\usepackage{amsthm}
\usepackage{algorithm}
\usepackage{algpseudocode}
\usepackage{hyperref}
\usepackage[bottom]{footmisc}

\theoremstyle{plain}
\newtheorem{thm}{Theorem}

\theoremstyle{definition}
\newtheorem{defn}[thm]{Definition} 

\begin{document}

\preprint{APS/123-QED}

\title[Frustrated Random Walks]{Frustrated Random Walks: A Faster Algorithm to Evaluate Node Distances on Connected and Undirected Graphs}
\author{Enzhi Li}
\email{enzhililsu@gmail.com}
\author{Zhengyi Le}
\affiliation {Suning R\&D Center, Palo Alto, USA}

\date{\today}

\begin{abstract}
Researchers have designed many algorithms to measure the distances between graph nodes, such as average hitting times of random walks, cosine distances from DeepWalk, personalized PageRank, etc. Successful although these algorithms are, still they are either underperforming or too time-consuming to be applicable to huge graphs that we encounter daily in this big data era. To address these issues, here we propose a faster algorithm based on an improved version of random walks that can beat DeepWalk results with more than ten times acceleration. The reason for this significant acceleration is that we can derive an analytical formula to calculate the expected hitting times of this random walk quickly. There is only one parameter (the power expansion order) in our algorithm, and the results are robust with respect to its changes. Therefore, we can directly find the optimal solution without fine-tuning of model parameters. Our method can be widely used for fraud detection, targeted ads, recommendation systems, topic-sensitive search, etc.

\end{abstract}
\maketitle

\section{Introduction}
\label{introduction}
Graph theory has been widely used to study social networks. A fundamental problem that commands researchers' attention in graph theory is how to quantitatively measure the distances between nodes. Once we have defined node distances, we can perform such actions as label propagation\cite{zoidi2015graph}, item recommendation\cite{fouss2007random}, and topic-sensitive search\cite{haveliwala2002topic, jeh2003scaling} in a social media that is represented as a graph. 

There are already many algorithms for calculating graph node distances, such as the geodesic distance from Dijkstra's algorithm, the cosine distance from DeepWalk\cite{perozzi2014deepwalk}/node2vec\cite{grover2016node2vec}, the expected hitting times of a random walk on a graph\cite{lovasz1993random, klein1993resistance}, and the distances from personalized PageRank algorithm\cite{page1999pagerank, haveliwala2002topic, jeh2003scaling}. The Dijkstra's algorithm, quick although it is, only considers the shortest path while discarding the rich structures of a graph. DeepWalk and node2vec algorithms, which consider global structures of graphs, can yield state-of-the-art results, yet the time and memory costs of running these algorithms are prohibitive. Random walk and personalized PageRank, which consume much fewer CPU resources, cannot, according to our findings, produce node ranking results that beat DeepWalk or node2vec method.

To address the above issues, here we propose a new algorithm that is based on an improved version of random walks on graphs to measure the distance between any pair of nodes in a graph using much shorter time and much less memory than DeepWalk. By calculating the node distances, we can gain a deeper and more intuitive understanding of the proximities of apparently disconnected social network users, since the proximity of a pair of users should be negatively correlated with their distance. With this algorithm, we can conveniently quantify the influence of a node upon another, using much fewer CPU hours. Here, by node distance, we mean how difficult it is to reach one node from another, and this distance function does not necessarily satisfy all the conditions imposed by mathematicians.  

In this paper, we introduce the Frustrated Random Walk (FRW) by augmenting a traditional random walk with an acceptance probability for each proposed transition, and use the expected hitting times of this frustrated random walk to measure node distances. We further derive an analytical formula to quickly calculate the expected hitting times. We show that the expected hitting times of frustrated random walks can yield node proximities that are in compliance with human judgement on a connected and undirected graph, whether be it weighed or unweighted. Using our method, we get results that beat DeepWalk algorithm, while using much fewer CPU hours. In summary, there are four advantages of our method compared with previous algorithms: 1. The distance function has an analytical expression; 2. The distance function possesses asymmetricity (see Section \ref{background} for details); 3. No parameter tuning is needed thanks to the stability of our results with respect to the only parameter in our method; 4. Our method is much faster than DeepWalk.

\section{Background}
\label{background}

Many algorithms are available for the calculation of graph node distances. Two distance functions that have gained much popularity are the geodesic distance\cite{baesens2015fraud} and the cosine distance which is a byproduct of DeepWalk\cite{perozzi2014deepwalk} or node2vec algorithm\cite{grover2016node2vec}. The geodesic distance between two nodes in an undirected graph is defined as the length of the shortest path connecting these two nodes. This distance can be quickly calculated using Dijkstra's algorithm, which, however, focuses only on the shortest path and ignores any alternative routes\cite{klein1993resistance}.
\begin{figure}[h!]
\centerline{\includegraphics[width = 0.75 \linewidth]{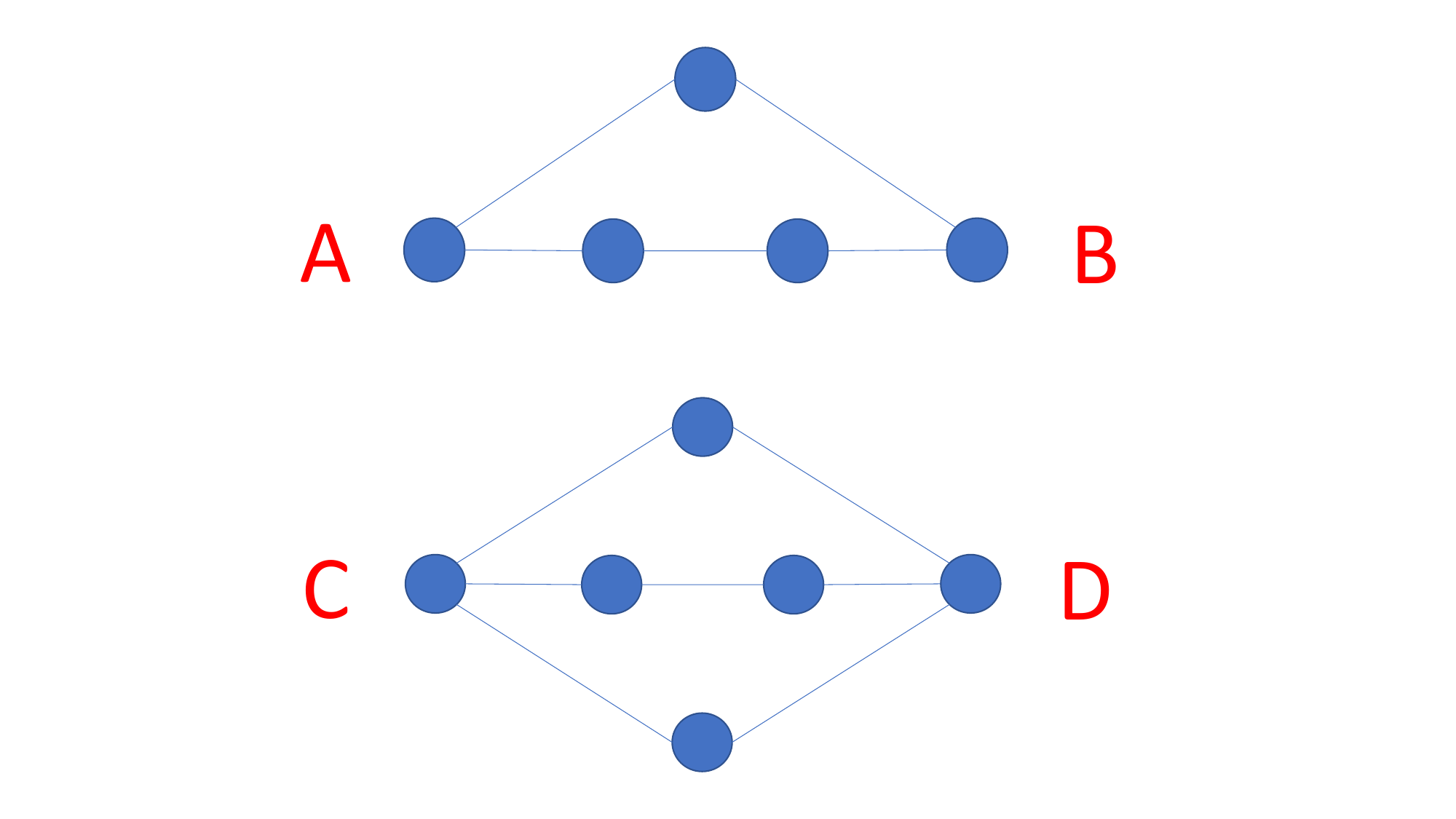}}
\caption{Geodesic distance v.s. hitting time expectation. $G_{AB}$: geodesic distance from $A$ to $B$; $H_{AB}$: expected hitting time from $A$ to $B$. $G_{AB} = G_{CD} = 2$, whereas $H_{AB} = 6 > H_{CD} = 5.25$. Intuitively, $H$ is a better metric than $G$. }
\label{graph_schema}
\end{figure}
Because of this, when we employ it to calculate the distance between two nodes in a graph, we have disregarded a significant amount of known information. The disregarding of the rich structures of a graph from which we could have extracted a cornucopia of precious information constitutes one major disadvantage of this algorithm. As shown in Fig. \ref{graph_schema}, we expect the distance between $C$ and $D$ to be shorter than that between $A$ and $B$ due to the availability of an extra path connecting the former pair. Numerical results show that expected hitting time as a measure of distance is more consistent with our intuition. 
 
 In DeepWalk and node2vec algorithms, we calculate the distance between two nodes by first mapping each node to a dense vector using the word2vec method\cite{mikolov2013distributed}, and then using the cosine distance between these two mapped dense vectors as graph node distance. This algorithm, which can be considered as an extension of the word2vec algorithm to graphs, requires the pre-existence of a node corpus that can be generated by performing tens of thousands of random walks on a graph. The generation of this corpus and the training of the model is pretty memory-intensive and time-consuming, thus precluding its application to extremely large graphs. 

\begin{figure}[h!]
\centerline{\includegraphics[width = 0.75 \linewidth]{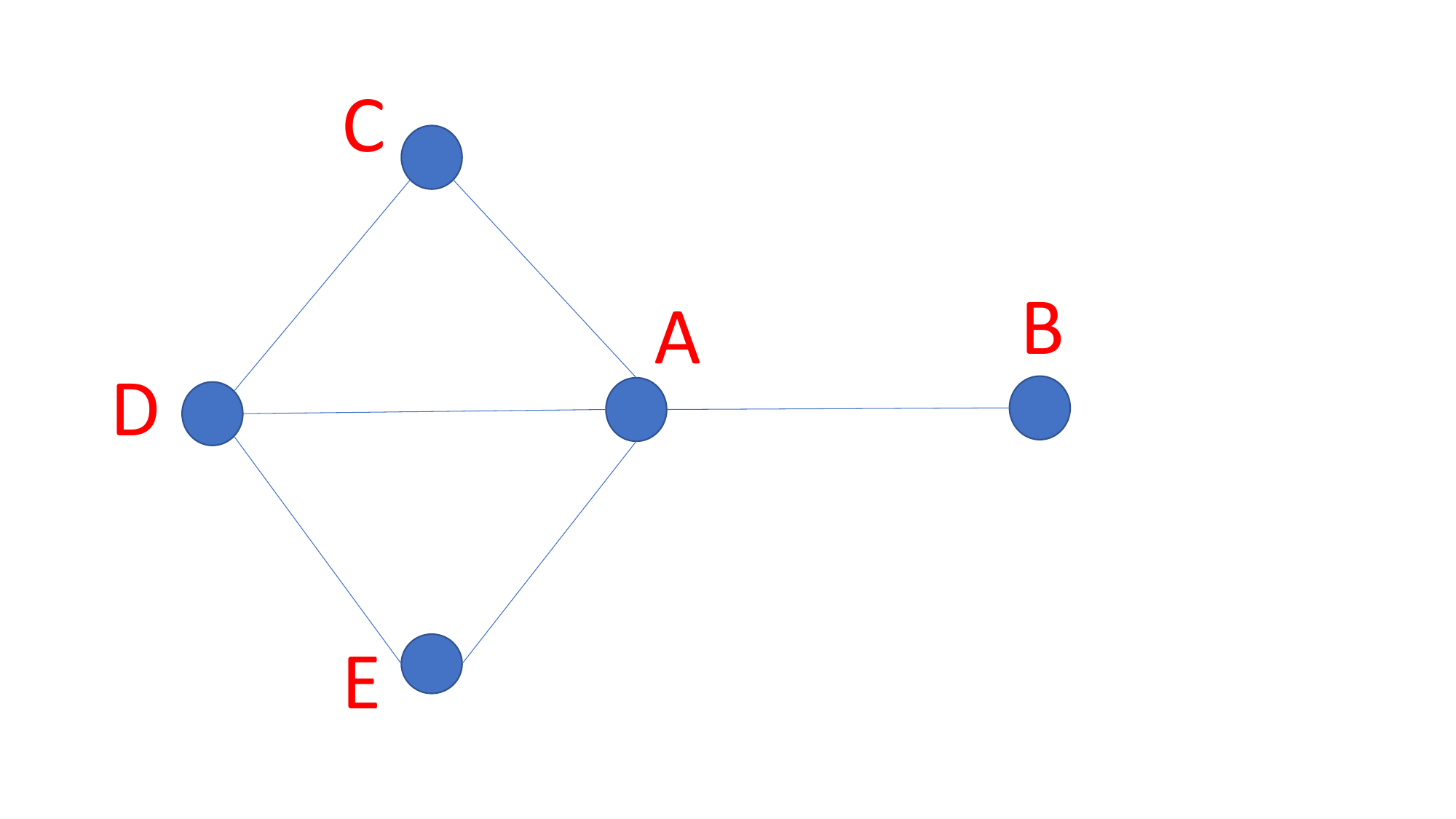}}
\caption{Asymmetricity of distances. $H_{AB}$: expected hitting time from $A$ to $B$. $H_{AB} = 11 \ne H_{BA} = 1$. Hitting time expectation as a distance function is asymmetric. Obviously, node $A$ is more closely associated with nodes $C, D, E$ than with $B$, we thus expect the distance from $A$ to $B$ should be longer than that from $B$ to $A$. }
\label{asymmetricity_figure}
\end{figure}

Another potential problem is that all these algorithms (geodesic distance, DeepWalk, and node2vec) yield symmetric distance functions for any pair of nodes in an undirected graph. The distance function is symmetric in the sense that the distance from node $A$ to $B$ is guaranteed to be identical to the distance in reversed direction. However, even for an undirected graph such as the friendship social network of Facebook, it is unreasonable to believe that the distance from an influential user to an obscure one should be invariant when the direction is reversed (see Fig. \ref{asymmetricity_figure} for an illustration). Since not all users of a social network have equal reputation and prestige, we claim that the relationships between social network users are non-equivalent, non-reflective, and asymmetric. Thus, for purpose of label propagation, a good definition of distance function between two nodes of a graph should consider this asymmetricity of relationships between nodes, even for undirected graphs where edge directions are meaningless or unavailable.

Here in this paper, we will concentrate our attention on random walks (See Section \ref{simple_random_walk_section} for a detailed description). There are many scenarios for performing random walks on graphs, one of which starts from a node, say $A$, and selects another node, say $B$, as its target, and performs a multitude of random walks starting from $A$ while counting how many times this random walker encounters node $B$ within a pre-specified number of steps\cite{fouss2007random}. The encountering frequency for a random walk provides a measure of the distance between nodes $A$ and $B$. The larger the frequency, the shorter the distance. This method of measuring the distance between two nodes, valid although it is, also has several drawbacks, the most prominent of which is its strong dependence on such capricious parameters as the maximum number of steps each random walker is allowed to traverse, and the number of random walks to be performed for the encountering frequency to be stable. Another weakness of this method is that it is again much too time-consuming to perform enough number of random walks to gain a statistically significant result for two nodes that are located afar in a colossal graph. The application of this random walk scenario to a small graph is no less problematic because a random walker starting from one node in a connected graph is guaranteed to reach any other node in the same graph as long as the random walk lasts long enough\cite{ross1996stochastic}, thus rendering all the distances between any pair of nodes almost the same. Given these considerations, we choose not to use the encountering frequency to measure node distances. Rather, we measure node distances using the expected hitting times of random walks, the details of which will be the theme of this paper. 

A convenient method to perform label propagation via random walks is to solve discrete Laplacian equations on graphs\cite{chung2000discrete, zhu2003semi}. Consider, for example, a graph in which we have labeled some nodes as "black", some as "white", and some as "unknown", as detailed in Ref. \cite{doyle1984random}. We can estimate the colors of the unknown nodes either by performing a Monte Carlo simulation or by solving a discrete Laplacian equation. The inference of these unknown colors is equivalent to performing label propagation in a graph. Ref. \cite{doyle1984random} shows that numerical solution of a discrete Laplacian equation gives us more accurate results while using far less time than Monte Carlo simulations. Solution of Laplacian equations requires the pre-specification of boundary conditions which are not always available\cite{zhu2003semi}. The black and white labels in Ref. \cite{doyle1984random} are the boundary conditions for a direct solution of the Laplacian equation to be feasible. However, if all the known labels are black, which is a common case in such real-world applications as fraud detection in a social network, then the only thing that a solution of the Laplacian equation can tell us is that all the colors of the unknown nodes should be black, which is practically useless to us. 

To overcome these difficulties, here we propose a new algorithm to measure node distances by invoking an exact solution to expected hitting times of frustrated random walks on a graph. A programmatic implementation of this algorithm can be found \href{https://github.com/suning-opensource/frustrated-random-walk}{here} \cite{program_implementation} . With the four advantages of our algorithm already summarized in Section \ref{introduction}, we will continue to present our algorithm in detail in the next section.

\section{Proposed method}
\label{method_outline}
In this section, we give an analytical method to calculate the expected hitting times of random walks on an undirected, connected and simple graph $G$. The connectedness of the graph is not a major restriction to our method due to the availability of efficient algorithms for finding connected components of an undirected graph. Because we are considering a social network of users who would have relationships only with others, we demand that the graph should be simple, meaning that none of the nodes are self-looped. 

The graph $G$ consists of vertices that constitute a set $\mathbb{V}$ and edges that connect these vertices. We denote an edge as $e_{ij}$ if it connects two vertices $i$ and $j$. The cardinality of set $\mathbb{V}$ is indicated by $|\mathbb{V}|$. Denote the adjacency matrix of this graph as $A$, whose dimension is $|\mathbb{V}| \times |\mathbb{V}|$, with matrix elements $A_{ij} = 1$ if there is an edge between node $i$ and node $j$, and $A_{ij} = 0$ otherwise. For a simple graph, we always have $A_{ii} = 0, \forall i \in \mathbb{V}$. The dimension of $A$ is the number of nodes in the graph, and the number of non-zero matrix elements is the edge number (or twice the edge number if all edges are undirected). $A$ is symmetric if graph $G$ is undirected, or else it is generally non-symmetric. We can associate weights to the edges and use $w_{ij}$ to denote the weight of edge $e_{ij}$. The degree of vertex $i$ is defined as $D_{i} = \sum_{j}w_{ij}$. 

In this paper, we calculate the probability of reaching a target node from any other node in the graph, whereas in a directed graph, a node may not be reachable from another node, and thus here we only consider undirected and connected graphs. For a random walker that starts from node $s$ and use node $t$ as its target, \textbf{the hitting time is defined as the number of steps the random walker needs to traverse before it reaches the target node $t$ for the first time}. Obviously, the hitting time from $s$ to $t$ is a random variable that depends on the graph structure, and we use $N^{(s)}_{t}$ to denote it. In this paper, we will use the expectation of hitting time to measure the distance from a starting node to a target node. Intuitively, the shorter the distance between any two nodes, the more related they are to each other. For this purpose, we are to calculate the hitting time probability distribution $P(N_t^{(s)} = n)$ and its expectation (first order moment) which is denoted as $ \langle N^{(s)}_{t} \rangle$. The calculation of the hitting probability and its expectation depends on the probability transition matrix $B$ which we will elaborate later. Our proposed method is an improvement of the traditional simple random walk algorithm. The major difference between the traditional algorithm and our algorithm is that in the former, a proposal to make a transition is always accepted, whereas in the latter, we have tried to frustrate the random walker by setting a threshold for each transition. We will describe these two algorithms in detail in the next sections and highlight their differences. 

\subsection{Review of simple random walk}
\label{simple_random_walk_section}
A simple random walk from a start node to a target node on a graph is described in Algorithm \ref{random_walk_algorithm}. 
\begin{algorithm}[H]
\caption{Simple random walk on a graph}
\label{random_walk_algorithm}
\begin{algorithmic}[1]
\Require{An undirected and connected Graph $G$}
\Procedure{RandomWalk}{$s, t$}\Comment{$s$ is the start node, and $t$ is the target node.}
	\State $c \gets s$
	\Repeat
		\State $r \gets $ a random neighbor of $c$
		\State $c \gets r$
	\Until{$c = t$}
\EndProcedure
\end{algorithmic}
\end{algorithm}

Assume that node $s$ has $m$ neighboring nodes, of which at most one is the target $t$. We enumerate these $m$ nodes using indices $i_s$. Then the probability $P(N_t^{(s)} = n)$ can be recursively represented as 
\begin{eqnarray}
\label{recursive_equation_simple_random_walk}
P(N_t^{(s)} = n) = \sum_{i_s \ne t} \frac{w_{s, i_s}}{D_s} P(N_t^{(i_s)} = n-1), n \ge 2
\end{eqnarray}
In the above equation, $i_s$ indicates one of the neighbors of node $s$, $D_s = \sum_{i_s} w_{s, i_s}$ is the total weight or degree of node $s$, $\frac{w_{s, i_s}}{D_s}$ represents the probability for the random walker to make a transition from node $s$ to node $i_s$, and $P(N_t^{(i_s)} = n-1)$ is the probability of reaching $t$ from $i_s$ after exactly $n - 1$ steps. 
By definition, we have $P(N_t^{(t)} = n) = 0, \forall n > 0$, which justifies our notation $i_s \ne t$ in the summation subscript in Eq. (\ref{recursive_equation_simple_random_walk}). If we scan all possible starting vertices $s$, we can obtain a simultaneous system of difference equations for the hitting probabilities $P(N_t^{(i)} = n), \forall i \in \mathbb{V}, i \ne t$. Specifically, if a vertex $j$ has only one single neighbor, and this very neighbor is just our target $t$, then we have $P(N_t^{(j)} = n) = \delta_{n, 1}$, with $\delta_{n, 1}$ being the Kronecker $\delta$ function. 
\begin{defn}
A node in graph $G$ is called an \textbf{\textit{adherent}} to a target if this node has the target as its only neighbor. 
\end{defn}
As an illustration, node $B$ in Fig. \ref{asymmetricity_figure} is an adherent if we use node $A$ as our target. 
Since we already know the probability distribution of hitting times of the adherents to target $t$, we can exclude those nodes when establishing difference equations for $P(N_t^{(s)} = n)$. With these preliminaries, we can write out a system of difference equations for $P(N_t^{(i)}= n)$: 
\begin{align}
\label{recursive_equation}
P(N_t^{(i)}= n) = \sum_{\substack{j = 1 \\ j \ne t}}^{|\mathbb{V}|} B_{ij} P(N_t^{(j)} = n - 1), n \ge 2, 
\end{align}
where we have imposed the restriction that the starting node $i$ should not be equal to $t$, and should not be an \textit{adherent} to the target. The $B$ matrix is called \textbf{probability transition matrix}, the elements of which are 

\begin{align}
B_{ij} = 
\begin{cases}
\frac{w_{ij}}{\sum_{j}w_{ij}} & A_{ij} \ne 0, \text{and } j \ne t; \\
0 & A_{ij} = 0 \text{, or } j = t.
\end{cases}
\end{align}

For most graphs that we encounter in the real world, $B$ is a sparse matrix. Note that although in an undirected graph the adjacency matrix $A$ is always symmetric, the probability transition matrix is generally non-symmetric. Moreover, due to the exclusion of the target node $t$ in the definition of the probability transition matrix, the sum of each row in $B$ is not necessarily equal to 1. The rule is that $\sum_{j}B_{ij} = 1$ if the target node $t$ is not a neighbor of node $i$; otherwise $\sum_{j}B_{ij} <1$. Since we have excluded the target node $t$ and all its adherents from the set of starting nodes, $B$ has a dimension that is smaller than the graph node number. 
For an undirected graph, $B$ is guaranteed to be square because a random walker starting from a node that is not the target cannot possibly reach an adherent node to the target. 
The fact that certain row sums of matrix $B$ are smaller than 1 means that this matrix is not a Markov matrix in the strictest sense (by definition, each row of a Markov matrix should sum to one), and that all of its eigenvalues have a magnitude that is smaller than 1 \cite{meyer2000matrix}. As a result, the spectral radius which is defined as the largest eigenvalue (magnitude) of matrix $B$ is also smaller than 1. We will take advantage of this fact later in this paper. 

\subsection{Frustrated random walk}
The procedure as described in Algorithm \ref{random_walk_algorithm} is the simplest random walk algorithm. It is used to generate node corpus for DeepWalk\cite{perozzi2014deepwalk}. The average hitting times of the simple random walks can also be used to measure graph node distances, see, for example, Ref. \cite{white2003algorithms, fouss2007random}. 

However, in many cases, direct application of this random walk procedure to the evaluation of node distances in a connected graph could give us misleading results, because this method tends to give undeserved advantages to the nodes which have only a few connections to the other nodes, yet all these few connections are directed towards some extremely prestigious nodes in the graph. For example, if we study the social relationships in \textit{Harry Potter} serials, we would expect Ron and Hermione to be closely associated with Harry Potter, who is the protagonist. However, when we measure the distances by calculating the average hitting times of a simple random walk, it turns out that Marge Dursley (sister in law of Harry Potter's auntie) is closest to Harry Potter. We can easily interpret this seemingly counter-intuitive result by noting that Marge Dursley appears rarely, yet almost all her appearances coincide with that of Harry Potter. As a result, by employing the simple random walk procedure, if we use the protagonist as our target, some minions could have shorter expected hitting times to the protagonist than the protagonist's close friends. 

\begin{figure}[h!]
\centerline{\includegraphics[width = 0.8 \linewidth]{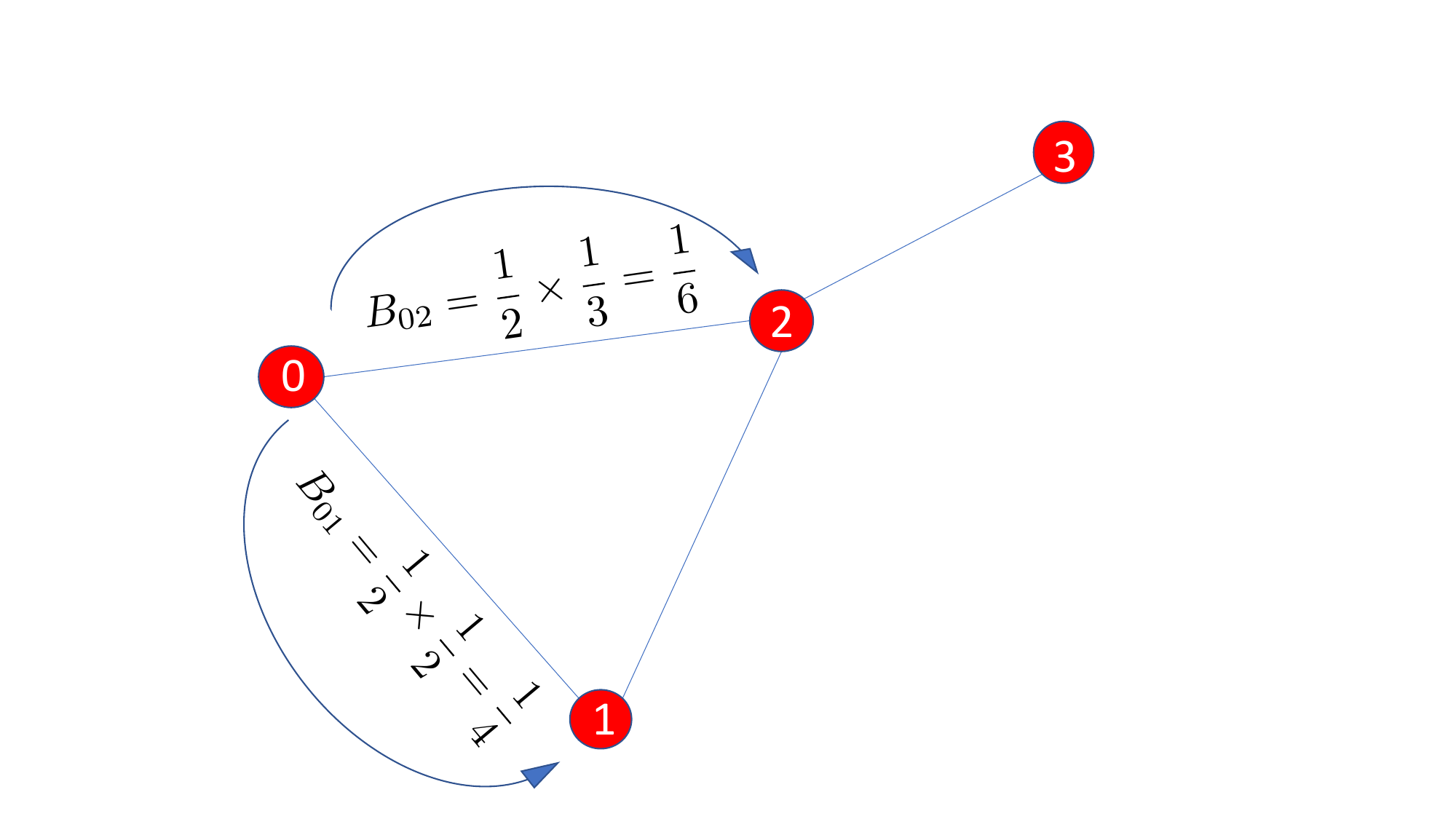}}
\caption{A schematic representation of frustrated random walks on a weightless undirected graph. Curves with arrows represent transitions.}
\label{frw_schema}
\end{figure}

To resolve the above issue which is inherent for simple random walks, here we propose an improved version of the random walk which we call the \textit{frustrated random walk}. In this new algorithm, we frustrate the random walker by introducing an acceptance probability to restrict the transition from a minion to the protagonist. This improved random walk is described in Algorithm \ref{frustrated_random_walk} and illustrated in Fig. \ref{frw_schema}.  

\begin{algorithm}[H]
\caption{Frustrated Random walk on a graph}
\label{frustrated_random_walk}
\begin{algorithmic}[1]
\Require{An undirected and connected Graph $G$}
\Procedure{FrustratedRandomWalk}{$s, t$}\Comment{$s$ is the start node, and $t$ is the target node.}
	\State $c \gets s$
	\Repeat
		\State $r \gets $ a random neighbor of $c$
		\State $c$ proposes to make a transition to $r$ with probability $P_{pro}$
		\State $r$ accepts the transition with probability $P_{acc}$
		\State $P_{transition}$ $\gets$ $P_{pro} \times P_{acc}$
		\State generate a random number $R \sim U(0, 1)$
		\If {$R < P_{transition}$}
			\State $c \gets r$
		\Else
			\State \textbf{continue}
		\EndIf
	\Until{$c = t$}
\EndProcedure
\end{algorithmic}
\end{algorithm}

The motivation for the introduction of frustration to transition is that a prestigious node in a graph may receive tens of thousands of proposals from its neighbors, and this node, being prim and arrogant, tend to accept proposals that are of higher values. Consequently, in our new algorithm, a minion would have a harder time attracting the attention of the hero, whereas the heroine would have a much better chance. Table \ref{harry_potter_transition_probabilities} illustrates the calculation of transition probabilities using simple and frustrated random walk methods for Harry Potter dataset (see Section \ref{comparison} for details). The transition probability of Hermione Granger to Harry Potter is higher (lower) than that from Marge Dursley for the frustrated random walk (simple random walk), thus justifying our introduction of acceptance probabilities in the frustrated random walk algorithm.

\begin{table}[]
\begin{tabular}{ccc}
\hline
 & SRW & FRW \\
 \hline
 Marge Dursley & $\frac{58}{172} = 0.337 $ & $\frac{58}{172} \times \frac{58}{36044} = 5.43 \times 10^{-4}$ \\
 Hermione Granger & $\frac{4539}{18053} = 0.251$ & $\frac{4539}{18053} \times \frac{4539}{36044} = 3.17 \times 10^{-2}$ \\
 \hline
\end{tabular}
\caption{Transition probabilities from Marge Dursley and Hermione Granger to Harry Potter using Simple Random Walk (SRW) and Frustrated Random Walk (FRW) procedures.}
\label{harry_potter_transition_probabilities}
\end{table}

We can calculate the average hitting time for a frustrated random walk following the methods outlined in the previous section, with the only modification being the probability transition matrix and the way we treat the adherents to the target. The off-diagonal matrix element $B_{ij}$, which represents the probability of hopping from vertex $i$ to a different vertex $j$, is now calculated as: 
\begin{align}
\label{frustrated_probability_transition_matrix}
B_{ij} = 
\begin{cases}
\frac{w_{ij}}{D_{i}} \frac{w_{ji}}{D_{j}} & A_{ij} \ne 0, \text{and } j \ne t; \\
0 & A_{ij} = 0 \text{, or } j = t. 
\end{cases}
\end{align}
Since we are only considering undirected graphs, we always have $w_{ij} = w_{ji}$. We can understand $w_{ij}/D_{i}$ as the probability of vertex $i$ proposing to make a transition to vertex $j$, and $w_{ji}/D_{j}$ as the probability of vertex $j$ accepting the transition from vertex $i$. Thus, we call $w_{ij}/D_{i}$ the \textbf{proposal probability}, $w_{ji}/D_{j}$ the \textbf{acceptance probability}, and ultimately, $B_{ij}$ the \textbf{transition probability}. 

In a simple random walk where the transition is always accepted, the random walker must jump to one of its neighbors for each step, and thus we have $B_{ii} = 0, \forall i \in \mathbb{V}$. On the other hand, in our frustrated random walk algorithm, the random walker's proposal to make a transition may be declined, in which case the walker will remain in its original position. For sake of convenience, we can understand this failed attempt to make a transition as a passive transition from a vertex to itself. Therefore, the diagonal elements of the probability transition matrix could be non-zero, which makes $B$ a primitive matrix according to the Perron-Frobenius theorem\cite{meyer2000matrix}. In order to calculate the diagonal elements of $B$, we observe that for a certain vertex $i$, if the target $t$ is not one of its neighbors, then $\sum_{j} B_{ij} = 1$, or else $\sum_{j, j \ne t}B_{ij} = 1 - \frac{w_{it}w_{ti}}{D_{i}D_{t}} < 1$. Thus, the diagonal elements of $B$ are
\begin{align}
\label{diagonal_matrix_elements}
B_{ii} = 
\begin{cases}
1 - \sum_{j, j \ne i} B_{ij}, & \text{if } t \text{ is not a neighbor of } i; \\
1 - \frac{w_{it}}{D_i}\frac{w_{ti}}{D_t} - \sum_{j, j \ne i} B_{ij}, & \text{if } t \text{ is a neighbor of } i. 
\end{cases}
\end{align}
We can conclude from this observation that the probability transition matrix is \textit{not} a Markov matrix, and that the spectral radius of this matrix is smaller than 1. We will take advantage of this fact later in this paper. 

Now that we have written out the transition matrix, we can continue to calculate the hitting time probability distribution. 
Following the methods detailed in Section \ref{simple_random_walk_section}, we can write a recursive equation for $P(N_{t}^{(i)} = n)$: 
\begin{align}
\label{frustrated_recursive_equation}
P(N_{t}^{(i)}= n) &= \sum_{j} B_{ij} P(N_t^{(j)} = n - 1) \\\nonumber
\end{align}
By definition, the probability for an \textit{adherent} to propose a transition to its neighbor is always equal to 1, whereas the acceptance probability for this proposal is $p = \frac{w}{D_{t}}$, where $w$ is the weight of the edge connecting target $t$ to the adherent, and $D_{t}$ is the degree of the target node. We can directly write out the probability of reaching the target after exactly $n$ steps from an adherent, which is 
\begin{align}
P(N_{t}^{adherent} = n) = (1 - p)^{n-1} p, n \ge 1, p = \frac{w}{D_{t}}
\end{align}
This is a geometric distribution, for which the expectation and variance are $\mu = 1/p$, and $\text{Var} = (1-p)/p^2$. 

\subsection{Analytical solution of expected hitting times}

Previously, we have obtained a recursive equation for $P(N_{t}^{(i)} = n), n \ge 2$, which is the probability of hitting target node $t$ starting from node $i$ after exactly $n$ steps. By introducing the notation $\boldsymbol{X}^{(i)}_{n} = P(N_{t}^{(i)} = n)$, we can rewrite Eq. (\ref{recursive_equation}) (for simple random walk) or Eq. (\ref{frustrated_recursive_equation}) (for frustrated random walk) into matrix form as 
\begin{align}
\label{matrix_equation}
\boldsymbol{X}_{n} = B \boldsymbol{X}_{n - 1}, n \ge 2
\end{align}
An iterative solution to Eq. (\ref{matrix_equation}) is 
\begin{align}
\label{iterative_solution}
\boldsymbol{X}_{n} = B^{n-1} \boldsymbol{X}_{1}, n \ge 1
\end{align}
The initial probability vector $\boldsymbol{X}_1$ can be conveniently obtained by the observation that $X_{1}^{(i)} = 0$ if $t$ is not a neighbor of node $i$. If $t$ is a neighbor of $i$, then $X_{1}^{(i)} = \frac{w_{it}}{D_{i}}$ for the simple random walk, and $X_{1}^{(i)} = \frac{w_{it}}{D_{i}}\frac{w_{ti}}{D_{t}}$ for the frustrated random walk. Now that we have obtained matrix $B$ and the initial probability vector $\boldsymbol{X}_{1}$, we can continue to calculate all the hitting probabilities for any valid starting node. 

Although we can calculate all the hitting probabilities, most of the time, we are more interested in the observable quantities of these probability distributions, i.e., the moments, which are
\begin{eqnarray}
\langle N_t^{(i)m} \rangle = \sum_{n = 1}^{\infty} P(N_t^{(i)} = n) n^{m} := \sum_{n = 1}^{\infty} X_{n}^{(i)} n^{m} 
\end{eqnarray}
The expectation and variance of the hitting time starting from any node $i$ can be calculated from the first and second order moments of the hitting time probability distribution. 

We can calculate all the moments of a distribution using probability generating function, which is defined as 
 
\begin{align}
\tilde{f}^{(i)}(z) = \sum_{n = 1}^{\infty} X_{n}^{(i)} z^{n}, |z| \le 1. 
\end{align}

To simplify notation, we define a vector-valued function $\boldsymbol{\tilde{f}}(z) $ as
\begin{align}
\label{probability_generating_function}
\boldsymbol{\tilde{f}}(z) = \sum_{n = 1}^{\infty} \boldsymbol{X}_{n} z^{n}
\end{align}
From the first and second order derivatives of $\boldsymbol{\tilde{f}}(z)$, we can calculate the first and second order moments of hitting times as follows:
\begin{align}
& \langle \boldsymbol{N}_{t} \rangle = \boldsymbol{\tilde{f}}^{\prime}(1) \\\nonumber
& \langle \boldsymbol{N}_t^2 \rangle = \boldsymbol{\tilde{f}}^{\prime\prime}(1) + \boldsymbol{\tilde{f}}^{\prime}(1) \\\nonumber
\end{align}
Plugging Eq. (\ref{iterative_solution}) into Eq. (\ref{probability_generating_function}) gives us 
\begin{align}
\label{generating_function}
\boldsymbol{\tilde{f}}(z) &= \Big( \sum_{n = 1}^{\infty} z^{n} B^{n-1} \Big) \boldsymbol{X}_{1} \\ \nonumber
&= z (I - z B)^{-1} \boldsymbol{X}_{1}
\end{align}
The second line of the above equation stems from the fact that $|z| \le 1$ and that the spectral radius of matrix $B$ is smaller than 1. 

Given the probability transition matrix $B$ and the initial probability vector $\boldsymbol{X}_{1}$, we can in principle calculate exactly the generating function and probability moments.  
However, inverting matrix $I - zB$ is no easy task, especially when the graph is huge. Moreover, the fact that the sparsity of matrix $B$ which we should take full advantage of can get lost after matrix inversion compels us to shun the idea of directly inverting $I - zB$. Instead, we try to calculate probability moments without inverting any matrix. For this purpose, we rewrite Eq. (\ref{generating_function}) as 
\begin{align}
\label{generating_function_reordered}
(I - z B) \boldsymbol{\tilde{f}}(z) = z \boldsymbol{X}_{1}
\end{align}

Differentiating the above equation with respect to $z$ on both sides and setting $z = 1$, we get 
\begin{align}
\label{first_order_moment_power_series}
\langle \boldsymbol{N}_{t} \rangle = \boldsymbol{\tilde{f}}^{\prime}(1) = \sum_{n = 0}^{\infty} B^{n} \boldsymbol{\tilde{f}}(1)
\end{align}
By definition, $\boldsymbol{\tilde{f}}(1) = \begin{pmatrix} 1 & 1 & 1 ... & 1\end{pmatrix}^{\text{T}}$. It is noteworthy that the first order moments are independent of the initial probability vector $\boldsymbol{X}_{1}$, and depend only on the transition matrix $B$. 
Higher order derivatives of Eq. (\ref{generating_function_reordered}) produce higher order moments. 

\subsection{Numerical computation of expected hitting times}
With the analytical formula Eq. (\ref{first_order_moment_power_series}) for calculating hitting time moments, we can implement these algorithms using numerical programs. 
The pseudocode for calculating the expected hitting times is shown in Algorithm \ref{hitting_time_pseudocode}. 

\begin{algorithm}[H]
\caption{Hitting time calculation algorithm}
\label{hitting_time}
\begin{algorithmic}[1]
\Require{An undirected and connected Graph $G$, and a target $t$}
\Require{Calculate probability transition matrix $B$ for a target $t$ using Eq. (\ref{frustrated_probability_transition_matrix})}
\Require{Maximum iteration number $N$ must be positive}
\Require{Error limit $\epsilon$ must be positive}
\Procedure{HittingTimeCalculator}{$B, N, \epsilon$}
	\State $i \gets 0$
	\State $d \gets B.dimension$\Comment{Dimension of matrix B}
	\State $ones \gets \textit{vector of all 1's, shape = (d, 1)}$
	\State $power \gets ones$
	\State $\mu \gets ones$ \Comment{$\mu$: expectation of hitting times}
	\While{$i \le N$}
		\State $i \gets i + 1$
		\State $power \gets B * power$\Comment{Matrix multiplication}
		\State $\mu \gets \mu + power$
		\State $error \gets \text{norm of } power$
		\If{$error < \epsilon $}
			\State \textbf{break}
		\EndIf
	\EndWhile
	\State \Return $\mu$
\EndProcedure
\end{algorithmic}
\label{hitting_time_pseudocode}
\end{algorithm}

We can see from Eq. (\ref{frustrated_probability_transition_matrix}) that for the frustrated random walk, matrix $B$ is always symmetric for an undirected graph, meaning all its eigenvalues are real. By Perron-Frobenius theorem\cite{meyer2000matrix}, the spectral radius of matrix $B$ is exactly its largest eigenvalue $\lambda_{max}$. Since the spectral radius of $B$ is smaller than 1, the summation of power series in Eq. (\ref{first_order_moment_power_series}) is guaranteed to converge. The difference $1 - \lambda_{max}$, called spectral gap\cite{masuda2017random}, dictates how fast the summation in Eq. (\ref{first_order_moment_power_series}) converges. The larger the spectral gap, the faster the convergence. From the Perron-Frobenius eigenvalue theorem\cite{meyer2000matrix}, we can crudely estimate the spectral radius by the following formula:  
\begin{align}
\label{estimate_spectral_radius}
\lambda_{max} \approx \frac{\sum_{ij} B_{ij}}{N_{B}}, 
\end{align}
where $N_{B}$ is the order of the square matrix $B$. Crude although this approximation is, we find in reality that, Eq. (\ref{estimate_spectral_radius}) can give us a pretty good estimation, with relative errors being less than 1\% for the datasets we study (see Table \ref{relative_errors}). Due to the fact that the spectral radius is guaranteed to be smaller than 1, we can obtain numerical results for the expected hitting times with arbitrary precision by terminating the summation in Eq. (\ref{first_order_moment_power_series}) at a power that is high enough. In practice, we shall strive to strike a balance between numerical precision and computational time.  

\begin{table}[]
\begin{tabular}{cccc}
\hline
Dataset & IMDB & arXiv & Harry Potter \\
\hline
Relative error & 0.02\% & 0.02\% & 0.6\% \\
\hline
\end{tabular}
\caption{Relative errors of spectral radii obtained from Eq. (\ref{estimate_spectral_radius}) and exact calculation in various datasets that we will use in Section \ref{comparison}.}
\label{relative_errors}
\end{table}

\section{Experimental results and algorithm analysis}
\label{comparison}
\subsection{Comparison with existing methods on real-world datasets}
In this section, we use real-world datasets to test our algorithm against three well-established algorithms: DeepWalk (\textbf{DW}), Simple Random Walk (\textbf{SRW}) and Personalized PageRank (\textbf{PPR}). Since we are calculating the proximity of a node to a target in a graph, we can only benchmark our computational results with human judgement or with DeepWalk results due to lack of ground truth. 

We first use \href{https://www.kaggle.com/carolzhangdc/imdb-5000-movie-dataset}{IMDB 5000 movie dataset} \cite{imdb_data} as an example. From this dataset, we create an undirected graph in which each node represents an actor, and two actors are connected by an edge if they co-star at least one movie, with the edge weight counting the number of movies they co-star. In this way, we create an undirected and weighted graph, from which we can extract all its connected components. Without loss of much information, we only focus on the largest connected component, which contains 4626 actors and 12667 edges. From among these 4626 actors, we choose one actor as our target and rank the other actors to it based on their proximities with respect to the target. Using FRW, SRW and PPR, we get three ranking results. To quantitatively measure the quality of these rankings, we use the results of DeepWalk as our benchmark. Employing DeepWalk, we map each actor to a dense vector and calculate the cosine distances of the other actors to the target. The smaller the cosine distance, the more adjacent a node is to the target. In this way, we get another ranking of nodes with respect to the target. To quantitatively measure the quality of rankings using different methods, we calculate the Spearman correlation coefficients for each target, with DeepWalk results as our benchmark. Spearman correlation coefficient is a real number within range [-1, 1] that can quantify how similar two rankings are to each other. The larger the coefficient, the more similar two rankings are. We compare these ranking results in Table \ref{imdb_results}. From the table, we can see that FRW, SRW and PPR generate almost equal Spearman correlation coefficients. This is probably due to the fact that the IMDB dataset is lightly weighted: almost all (97.7\%) of the edges in this dataset have unit weight, and that the maximum edge weight of this dataset is just 6 (only two edges with such a weight). We notice that for a lightly weighted graph, FRW, SRW, PPR and DeepWalk yield similar ranking results. 

\begin{table*}
\begin{tabular}{|c|c|c|c|c|c|c|}
\hline
target & Robert De Niro & Morgan Freeman & Nicolas Cage & Leonardo DiCaprio & Jennifer Lawrence & Casey Affleck\\
\hline
PageRank value & 0.002891 & 0.002379 & 0.001993 & 0.001261 & 0.000907 & 5.8e-05\\
FRW & 0.4501 & 0.2941 & 0.4676 & 0.4068 & \textbf{0.4722} & 0.3808\\
SRW & 0.3981 & 0.2558 & 0.3494 & 0.3855 & 0.367 & 0.5107\\
PPR & \textbf{0.5359} & \textbf{0.4121} & \textbf{0.5228} & \textbf{0.5205} & 0.4598 & \textbf{0.5477}\\
\hline
\end{tabular}
\caption{Comparison of ranking results using DeepWalk as a benchmark. The first row lists all the targets with decreasing PageRank values from left to right, the second row gives the PageRank value of each target, and the following three lines list Spearman correlation coefficients between FRW, SRW, and PPR ranking results and DeepWalk ranking results, respectively. A large correlation coefficient indicates a high similarity between ranking results. }
\label{imdb_results}
\end{table*}

We next study a co-authorship network that we create by analyzing articles published in arXiv under the category "Computer Science $>$ Social and Information Networks" from 2010 to 2019\cite{real_world_graph_data}. Each author in the network is represented as a node, and two authors are connected by a weighted and undirected edge if they co-author at least one article, with edge weight counting the number of articles they co-author. From this network, we extract a connected component that contains 10,957 authors and 35,491 edges. We analyze this component using both DeepWalk and FRW algorithms following the procedures described above, and list some ranking results in Table \ref{arxiv_results}. From the table, we can see that the results from DeepWalk, FRW and PPR are consistent with each other, whereas the Spearman correlation coefficients between SRW and DeepWalk ranking results decays as PageRank value of target node decreases. This arises from the fact that although the arXiv dataset is more heavily weighted than IMDB dataset, it is still pretty lightly weighted compare to the dataset we are going study later. In our arXiv dataset, $84.36\%$ of the edges have weight 1, and the maximum weight of edges is 23. For a mildly weighted dataset such as the arXiv dataset, ranking results DeepWalk deviate from SRW significantly when PageRank value of the target is small, while FRW and PPR are still performing well. 

\begin{table*}
\begin{tabular}{|c|c|c|c|c|c|c|c|}
\hline
target & Jure Leskovec & Emilio Ferrara & Mason A. Porter & M. E. J. Newman & Stephen Boyd & Devon Callahan & Elaheh Raisi\\
\hline
PageRank value & 0.001801 & 0.001651 & 0.001575 & 0.000443 & 0.000127 & 7.7e-05 & 1.8e-05\\
FRW & \textbf{0.6607} & 0.3228 & 0.3179 & 0.4263 & \textbf{0.6431} & \textbf{0.3084} & 0.4383\\
SRW & 0.6514 & \textbf{0.6392} & 0.4916 & 0.3241 & 0.1924 & -0.0989 & 0.0092\\
PPR & 0.6126 & \textbf{0.6392} & \textbf{0.6313} & \textbf{0.5258} & 0.6013 & 0.2977 & \textbf{0.4434}\\
\hline
\end{tabular}
\caption{Comparison of ranking results for different targets. See the caption of Table \ref{imdb_results} for table interpretation.}
\label{arxiv_results}
\end{table*}

Our algorithm originates from the observation of how people tend to interact with each other in the real world, and thus we expect our algorithm to reach its peak performance in the dataset that can accurately capture authentic human-human interactions. We demonstrate this by analyzing a \href{https://github.com/PrimerLi/graph-data}{Harry Potter dataset} that we create from the \textit{Harry Potter} novels written by J. K. Rowling\cite{real_world_graph_data}. For this purpose, we represent each character in the novel as a node in a graph. If two characters co-appear in the same scene, then they are linked by an edge, the weight of which counts their co-appearance times. In total, there are 183 nodes and 4553 edges in this dataset. What differentiates this dataset from the previous two datasets is that the edge weights for this one range from 5280 to 1, whereas the edge weights range from 6 to 1 (29 to 1) for IMDB dataset (arXiv dataset). We thus call the Harry Potter dataset \textit{heavily weighted}, and the other two datasets \textit{lightly weighted}. We apply FRW, SRW and PPR algorithms to the Harry Potter dataset, and calculate Spearman correlation coefficients between the ranking results of these algorithms and that of DeepWalk, and tabulate the results in Table \ref{harry_potter_results}. From the table, we see that FRW almost always has the largest Spearman correlation coefficients no matter what target we have chosen. On the other hand, PPR has large coefficients when the target node has a large PageRank value, and SRW has large coefficients when the target has a small PageRank value.  

\begin{table*}
\begin{tabular}{|c|c|c|c|c|c|c|c|}
\hline
target & Harry Potter & Hermione Granger & Voldemort & Peter Pettigrew & Lucius Malfoy & Nagini & Charity Burbage\\
\hline
PageRank value & 0.162442 & 0.079422 & 0.022098 & 0.005564 & 0.003726 & 0.0014 & 0.000922\\
FRW & \textbf{0.7769} & \textbf{0.6999} & 0.4942 & \textbf{0.6633} & 0.5016 & \textbf{0.5764} & 0.5078\\
SRW & 0.3556 & 0.0001 & 0.0892 & 0.4291 & 0.5068 & 0.5561 & \textbf{0.6093}\\
PPR & 0.703 & 0.6815 & \textbf{0.7139} & 0.6192 & \textbf{0.602} & 0.4829 & 0.0182\\
\hline
\end{tabular}
\caption{Comparison of ranking results for different targets of Harry Potter dataset. See the caption of Table \ref{imdb_results} for table interpretation.}
\label{harry_potter_results}
\end{table*}

In order to see in detail how these ranking algorithms perform for the Harry Potter dataset, we list the top five most adjacent neighbors to targets of high and low PageRank values. In Tables \ref{Harry_Potter_high_PageRank} and \ref{Harry_Potter_low_PageRank}, we show the results from Harry Potter dataset. When we use a node (Harry Potter) with high PageRank value as our target, the SRW results deviate from the other three results significantly, whereas for a target (Lucius Malfoy) with low PageRank value, PPR deviates from the other three noticeably. We conclude that PPR biases favorably towards the central nodes in a graph, i.e., the higher a node’s PageRank value, the more likely it is to be considered adjacent to a target, whatever the target is. This could undermine PPR’s power for topic-sensitive search. From our knowledge of \textit{Harry Potter} novels, we claim that DeepWalk and FRW surpass the other two methods for the dataset we tested. 

\begin{table*}
\begin{tabular}{|c|c|c|c|}
\hline
DW & FRW & SRW & PPR\\
\hline
Ron Weasley & Ron Weasley & Marge Dursley & Ron Weasley\\
Hermione Granger & Hermione Granger & Dudley Dursley & Hermione Granger\\
Albus Dumbledore & Severus Snape & Vernon Dursley & Albus Dumbledore\\
Severus Snape & Dobby & Dobby & Rubeus Hagrid\\
Ginny Weasley & Sirius Black & Petunia Dursley & Severus Snape\\
\hline
\end{tabular}
\caption{Comparison of top nearest neighbors of Harry Potter, whose PageRank value is 0.1624 (rank order is 1/183). }
\label{Harry_Potter_high_PageRank}
\end{table*}

\begin{table*}
\begin{tabular}{|c|c|c|c|}
\hline
DW & FRW & SRW & PPR\\
\hline
Bellatrix Lestrange & Narcissa Malfoy & Narcissa Malfoy & Harry Potter\\
Narcissa Malfoy & Bellatrix Lestrange & Rabastan Lestrange & Ron Weasley\\
Severus Snape & Fenrir Greyback & Rodolphus Lestrange & Hermione Granger\\
Harry Potter & Scabior & Charity Burbage & Albus Dumbledore\\
Ron Weasley & Draco Malfoy & Andromeda Tonks & Draco Malfoy\\
\hline
\end{tabular}
\caption{Comparison of top nearest neighbors of Lucius Malfoy, whose PageRank value is 0.0037(rank order is 50/183). }
\label{Harry_Potter_low_PageRank}
\end{table*}

To further see how our algorithm outperforms both SRW and PPR for heavily weighted datasets that can reflect genuine human-human interactions, we continue to use another novel with highly involved human relationships as our test dataset. We extract a \href{https://github.com/PrimerLi/red_chamber_dream_network/blob/master/edges_pinyin.csv}{graph dataset} from \textit{Dream of the Red Chamber} written by Cao Xueqin, and apply the four algorithms (FRW, SRW, PPR, and DeepWalk) to it\cite{red_chamber_dream_data}. There are 324 vertices and 5807 edges in this graph. This dataset is heavily weighted in that the edge weights of this graph range from 1 to 418. We still use DeepWalk rankings as the benchmark, and calculate the Spearman correlation coefficients for each FRW, SRW and PPR algorithm, as shown in Table \ref{red_chamber_dream_results}. Similar to the Harry Potter dataset, here we also see that SRW deviates from DeepWalk when the target PageRank value is high, and PPR deviates from DeepWalk when target PageRank value is low. For all the targets we consider, FRW is on par with DeepWalk. We also study in detail (not shown) the top adjacent neighbors to each target, and find that FRW and DeepWalk indeed yield results that are consistent with human judgement. 

\begin{table*}
\begin{tabular}{|c|c|c|c|c|c|c|c|c|c|}
\hline
target & Jia Baoyu & Wang Xifeng & Lin Daiyu & Jia Zheng & Jia Zhen & Jia Yucun & Zhen Shiyin & Liu Xianglian & Wang Yitie\\
\hline
PageRank value & 0.062469 & 0.04258 & 0.032765 & 0.019667 & 0.016038 & 0.004841 & 0.002431 & 0.001409 & 0.00051\\
FRW & \textbf{0.6523} & 0.7174 & 0.3692 & 0.6015 & 0.5083 & \textbf{0.6297} & 0.4787 & 0.5242 & 0.5747\\
SRW & -0.1028 & 0.1275 & 0.2002 & 0.2725 & 0.4062 & 0.584 & \textbf{0.646} & \textbf{0.5288} & \textbf{0.6908}\\
PPR & 0.6043 & \textbf{0.7347} & \textbf{0.4808} & \textbf{0.6025} & \textbf{0.6198} & 0.1988 & 0.2748 & 0.0685 & -0.2208\\
\hline
\end{tabular}
\caption{Comparison of ranking results for different targets in \textit{Dream of the Red Chamber} dataset. See the caption of Table \ref{imdb_results} for table interpretation.}
\label{red_chamber_dream_results}
\end{table*}

\begin{figure}[h!]
\centerline{\includegraphics[width = \linewidth]{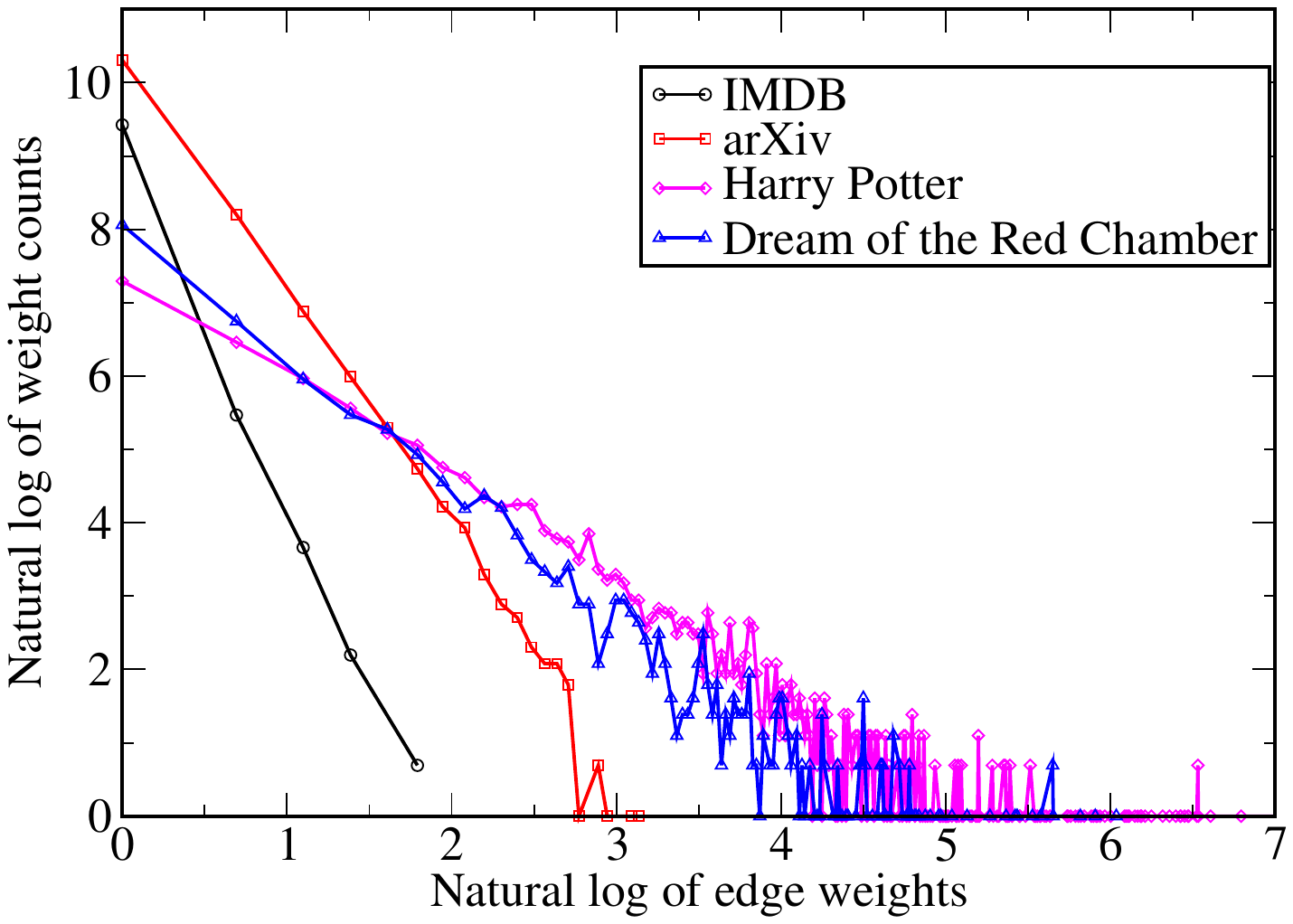}}
\caption{$\log-\log$ plot of edge weight distributions of lightly (IMDB and arXiv) and heavily (Harry Potter and \text{Dream of the Red Chamber}) weighted datasets. }
\label{edge_weight_distribution}
\end{figure}

We have noted that the FRW algorithm outperforms the competing algorithms only on heavily weighted datasets. For the lightly weighted datasets, the relatively simpler SRW and PPR methods suffice to deliver results of satisfactory quality. However, for heavily weighted graphs, SRW is beset by the problem of giving undeserved favor to the minions of the protagonists, and PPR tends to bias favorably towards the central nodes in a graph, whatever the target node is. To highlight the differences between lightly weighted and heavily weighted datasets, we have plotted the $\log-\log$ edge weight distributions in Fig. \ref{edge_weight_distribution}. Since the real-world human relations are generally heavily weighted, we expect the FRW algorithm to play a major role in future applications. 

\subsection{Benchmark frustrated random walk with other deep-learning-based graph algorithms}
In the wake of DeepWalk, people have proposed many ensuing algorithms for graph node embedding that are based on deep learning or graph convolutional networks\cite{tang2015line, wang2016structural, hamilton2017inductive, hamilton2017representation, abu2018watch}. In order to fully capture the advancements in the field of graph node embedding, here we benchmark the frustrated random walk results with these emerging techniques. We have shown in the previous sub-section that frustrated random walk (FRW) and DeepWalk outperform the other traditional non-deep learning methods for heavily weighted graphs (the \textit{Harry Potter} and \textit{Dream of the Red Chamber} datasets), and here in this section we will focus our attention on the \textit{Dream of the Red Chamber} dataset, and still use the DeepWalk results as our ground truth label because its results are mostly compatible with human judgement. 

Hopefully without loss of generality, we employ large-scale information network embedding (LINE)\cite{tang2015line}, structural deep network embedding (SDNE)\cite{wang2016structural}, and graph attention node embedding (GANE)\cite{abu2018watch} as our baseline algorithms. We choose to leave out graphSAGE \cite{hamilton2017inductive} algorithm because it requires that graph nodes should possess non-trivial features, whereas the frustrated random walk algorithm is designed only for graphs with featureless nodes. If all the nodes are featureless, then graphSAGE can be fully replaced with DeepWalk or node2vec which we have already considered in the last sub-section. 

By applying these algorithms to the \textit{Dream of the Red Chamber} dataset, we obtain ranking of nodes relative to targets and calculate Spearman correlation coefficients for each method with respect to DeepWalk results, as shown in Table \ref{representation_learning}. For sake of clarity, we have chosen the same set of target nodes as we used in the last sub-section. We have endeavored to make sure that the targets with a wide range of PageRank values can represent the full spectrum of nodes in the dataset, whether be it central or marginal. As is clear from the table, graph attention node embedding (with four maximum coefficients) and frustrated random walk (with three maximum coefficients) algorithms deliver results that are most consistent with that of DeepWalk. We thus conclude that we can beat the highly complicated deep-learning-based graph convolutional networks with our frustrated random walk algorithm for the task of node ranking, using much fewer computing resources, at least for the dataset we are studying. 

\begin{table*}
\begin{tabular}{|c|c|c|c|c|c|c|c|c|c|}
\hline
target & Jia Baoyu & Wang Xifeng & Lin Daiyu & Jia Zheng & Jia Zhen & Jia Yucun & Zhen Shiyin & Liu Xianglian & Wang Yitie\\
\hline
PageRank value & 0.062469 & 0.04258 & 0.032765 & 0.019667 & 0.016038 & 0.004841 & 0.002431 & 0.001409 & 0.00051\\
FRW & 0.6523 & 0.7174 & 0.3692 & \textbf{0.6015} & 0.5083 & \textbf{0.6297} & 0.4787 & \textbf{0.5242} & 0.5747\\
LINE & 0.2305 & 0.3124 & \textbf{0.5062} & 0.2681 & 0.1406 & 0.2138 & 0.4946 & 0.3338 & 0.6103\\
SDNE & 0.3396 & 0.1018 & 0.13 & 0.1144 & 0.1434 & 0.1719 & \textbf{0.6238} & 0.2277 & 0.5563\\
GANE & \textbf{0.6598} & \textbf{0.7389} & 0.461 & 0.4733 & \textbf{0.532} & 0.4105 & 0.5393 & 0.4052 & \textbf{0.6999}\\
\hline
\end{tabular}
\caption{Comparison of relative ranking results obtained from FRW and three representation learning algorithms, still using DeepWalk as benchmark. For each target node and each method, Spearman correlation coefficients with respect to DeepWalk ranking results are calculated and displayed in the corresponding cell in the above table. }
\label{representation_learning}
\end{table*}

\subsection{Parameter sensitivity}
\label{parameter_sensitivity}
\begin{figure}[h!]
\centerline{\includegraphics[width = \linewidth]{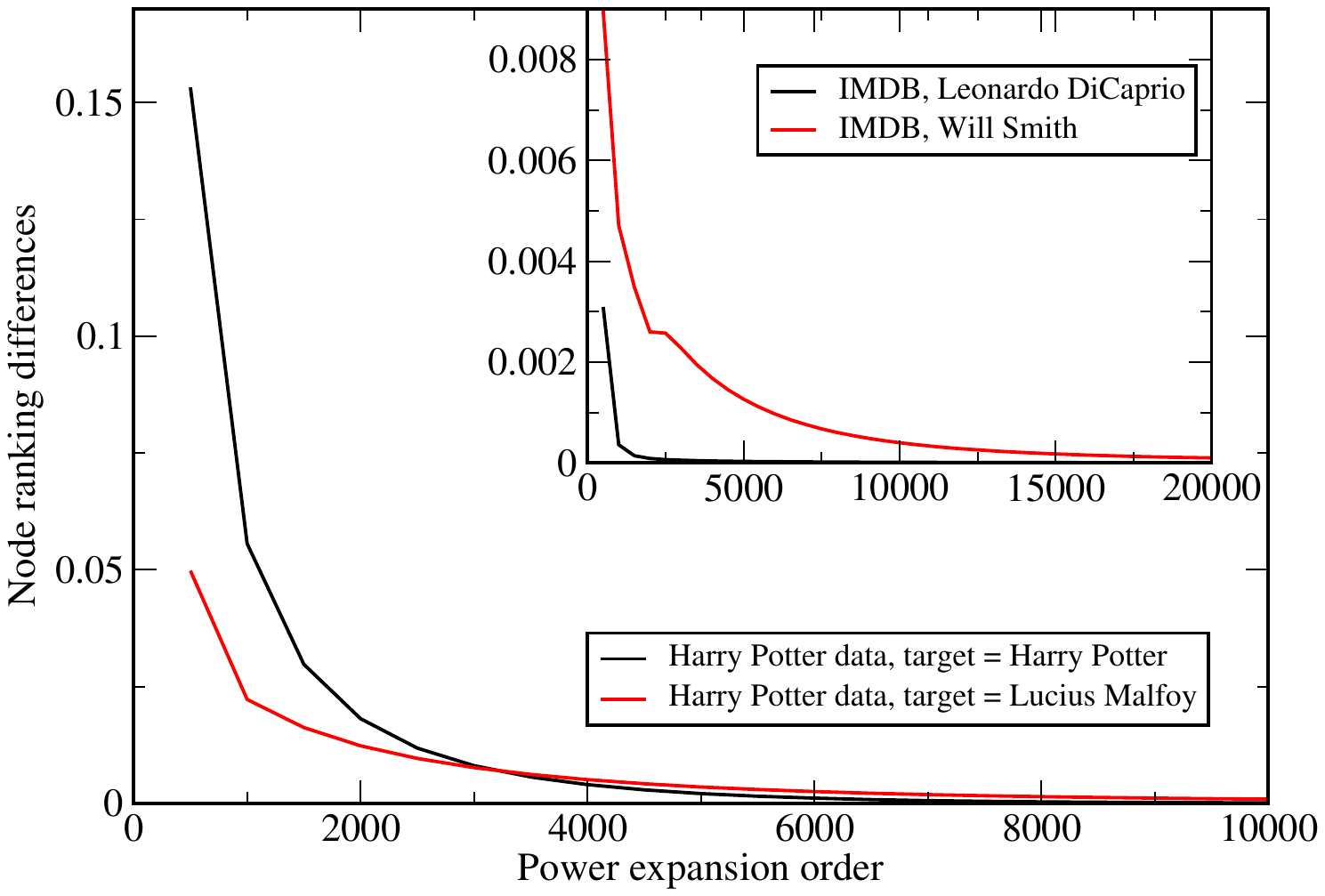}}
\caption{Ranking differences vs. power expansion order $N_{max}$ for Harry Potter dataset (shown in the main panel, with Harry Potter and Lucius Malfoy as targets) and IMDB dataset (shown in inset, with Leonardo DiCaprio and Will Smith as targets). All the curves decay to zero rapidly, indicating the robustness of the relative node ranking with respect to $N_{max}$ in our algorithm.
}
\label{counter_diff}
\end{figure}

During our experiments, we notice that although the average hitting time depends sensitively on the cut-off $N_{max}$ in the power series expansion $\langle \boldsymbol{N}_{t} \rangle = \sum_{n = 0}^{N_{max}} B^{n} \mathbf{1}$, the ranking of the nodes' proximities relative to a target is pretty stable. In Fig. \ref{counter_diff}, we plot the ranking differences vs. expansion order and find that all the curves decay to zero exponentially fast. The ranking difference, which is calculated by subtracting current normalized node similarity results (maximum similarity scaled to 1, and minimum similarity scaled to 0) from the previous one, equals zero when the node ranking stabilizes with respect to variations of $N_{max}$. Since $N_{max}$ is the only parameter in our method, we do not have to tune any parameter carefully to obtain optimal results, which renders our algorithm advantageous over other deep-learning-based methods such as DeepWalk, node2vec and the methods listed in the previous sub-section which are beset with a bunch of hard-to-select hyper-parameters. 

\subsection{Comparison of running speeds}
As we have noted above, DeepWalk and Frustrated Random Walk (FRW) can deliver the most competitive results that are consistent with our intuition. Here, we compare the running times of these two algorithms and show that FRW can beat DeepWalk in speed. We run our programs on Linux CentOS-7, with 32 cores and 128 Gigabyte memory. For FRW, which has only one parameter $N_{max}$, we terminate the power series in Eq. (\ref{first_order_moment_power_series}) when the node ranking difference (defined in Section \ref{parameter_sensitivity}) falls below a threshold value $10^{-4}$. The running time of FRW depends on the target we select, and in Table \ref{timing_comparison} we show the mean and standard deviation of running times for FRW with randomly selected targets (89 targets for IMDB, 82 targets for arXiv, and 99 targets for Harry Potter).  

When running the DeepWalk program, we first create a node corpus that contains 10,000 random paths, with the length of each path being 800. The generation of these random paths can be easily parallelized. We use the skip-gram method (the window size is set to be 5, and the minimum count is 10) implemented in Gensim\cite{rehurek_lrec} to train the model and map each node to a vector of size 300. The model training program of Gensim is already accelerated by multiprocessing parallelization, and we train the model using 32 processes. 

All the running time results are shown in Table \ref{timing_comparison}. From the table, we can see that for the specific task of node ranking relative to a target, we have achieved significant accelerations ($7 - 34$ times speedup) by switching from DeepWalk to FRW. It is also seen that the larger the dataset, the more significant the acceleration, which prompts us to adopt FRW for the large graphs that we encounter daily. 
 
\begin{table}
\begin{tabular}{cccc}
\hline
Dataset & IMDB & arXiv & Harry Potter \\
\hline
FRW (seconds) & $8.45 \pm 7.32$ & $21.5 \pm 34.9$ & $10.85 \pm 4.39 $\\
DeepWalk (seconds) & 191.2 & 742.4 & 76.5 \\
\hline
\end{tabular}
\caption{Running times of FRW and DeepWalk for different datasets.}
\label{timing_comparison}
\end{table}

\subsection{Time complexity analysis}
As we can see from Eq. (\ref{first_order_moment_power_series}) and Algorithm \ref{hitting_time}, the running time of our program depends on two factors: the graph structure and the spectral radius of probability transition matrix $B$. There are two loops in Algorithm \ref{hitting_time}. One is the outer loop whose number is determined by the power expansion order $N_{max}$, and another is a hidden loop implied in the sparse matrix-vector product. From Eq. (\ref{first_order_moment_power_series}), we know that in order for the power series to converge, we need to compute the series at least up to order $N_{max}$ the value of which is dictated by the spectral radius $\lambda_{max}$ of transition matrix $B$. We have already shown that the spectral radius is guaranteed to be smaller than one, and thus the series always converges. The nearer the spectral radius is to one, the slower the convergence rate is. From Eq. (\ref{estimate_spectral_radius}), we can estimate the expansion order $N_{max}$ as 
\begin{align}
N_{max} \propto \frac{1}{1 - \lambda_{max}} \approx \frac{N_{B}}{N_{B} - \sum_{ij}B_{ij}}. 
\end{align}
In the above equation, $N_{B}$ is the dimension of matrix $B$. From Eq. (\ref{diagonal_matrix_elements}), we see that not all row sums of $B$ are equal to one. We can rewrite Eq. (\ref{diagonal_matrix_elements}) to calculate the row sums as 
\begin{align}
\sum_{j}B_{ij} = 
\begin{cases}
1, & \text{if } t \text{ is not a neighbor of } i; \\
1 - \frac{w_{it}}{D_i}\frac{w_{ti}}{D_t}, & \text{if } t \text{ is a neighbor of } i. 
\end{cases}
\end{align}
The sum of all matrix elements of $B$ is thus 
\begin{align}
\sum_{ij}B_{ij} = N_{B} - \sum_{i \in \{\tilde{t}\}}  \frac{w_{it}}{D_{i}} \frac{w_{ti}}{D_{t}},
\end{align}
where, we have used the notation $ \{\tilde{t}\} := \{\text{neighbors of } t\}$. A naive approximation of the sum over neighbors of target is 
\begin{align}
\sum_{i \in \{\tilde{t}\}}  \frac{w_{it}}{D_{i}} \frac{w_{ti}}{D_{t}} \approx 1
\end{align}
Therefore, we have arrived at an estimation of the power expansion order $N_{max}$, which is 
\begin{align}
N_{max} \approx N_{B} \lesssim V
\end{align}
Here, for sake of simplicity, we have used $V$ to denote the number of vertices in the graph. 
Having estimated the order of the outer loop in Algorithm \ref{hitting_time}, we can continue to evaluate the order of the inner loop, which is the product of sparse matrix $B$ and a dense vector of size $N_{B}$. We can represent the inner loop as 
\begin{align}
v^{\prime}_{i} = \sum_{j}B_{ij}v_{j}, i = 0, 1, 2, ..., N_{B} - 1
\end{align}
From Eq. (\ref{frustrated_probability_transition_matrix}) and Eq. (\ref{diagonal_matrix_elements}), we see that the number of operations for calculating each $v^{\prime}_{i}$ is $D_{i} + 1$, where $D_{i}$ counts the number of edges connecting vertex $i$ to its neighbors. If we scan all possible values of $i$, then the total number of operations in the inner loop is $\sum_{i}\Big( D_{i} + 1\Big) \lesssim 2E + V$, where $E$ is the total number of edges in the graph. Multiplying together the outer loop number and the inner loop number, we can finally estimate the total number of operations in the program as
\begin{align}
N_{total} \approx (2E + V) V
\end{align}
For a sparse graph which is a common case in most real world applications, $E \propto V$, and thus the time complexity of our algorithm is $O(V^2)$.

\section{Conclusion}
\label{conclusion}
In this paper, we have proposed a new algorithm for performing random walks on graphs and derived an analytical formula to calculate its expected hitting times. This formula is valid for both simple random walk and frustrated random walk. The derivation and application of the formula only rely on the probability transition matrix that can be conveniently extracted from the graph's adjacency matrix. We also propose a quick method to numerically implement this formula, without the need to invert a possibly huge sparse matrix. The advantage of the analytical method over Monte Carlo simulations is that the former is much quicker and more accurate than the latter.  

We tested our algorithm against the DeepWalk method, and showed for some real-world datasets that our algorithm can produce node ranking results that are consistent with human judgement and that of DeepWalk. We also applied simple random walk and personalized PageRank to the same datasets and found that neither of these two methods can compete with DeepWalk or frustrated random walk algorithm for complicated graphs. We further compared our method with more recently proposed deep-learning-based algorithms, and found that the frustrated random walk method can compete with or outperform all of these algorithms. 

We studied model parameter sensitivity and found that if we only concern ourselves with nodes' relative rankings, then the program results are pretty stable with respect to variations of power series expansion order, which is the only parameter in our model. As a result, no parameter tuning is required for our method to deliver competitive results. For the task of finding the most adjacent nodes to a target node in a graph, frustrated random walk outperforms DeepWalk with a large margin in running speed. The time complexity of our algorithm is $O((E+V)V)$, and for sparse graphs, this is approximately $O(V^2)$. 

\begin{acknowledgements}
We would like to thank Suning R \& D Center for providing us with support and computing resources, without which this paper could not have been finished.  
\end{acknowledgements}

\bibliography{ref}
\bibliographystyle{apsrev}

\appendix 
\section{Experimental evidence on artificial dataset}
\label{artificial_dataset}
\subsection{An analytical calculation of hitting time distribution on a simple graph}
\label{exact_solution}
\begin{figure}[h!]
\centerline{\includegraphics[width= 0.8 \linewidth]{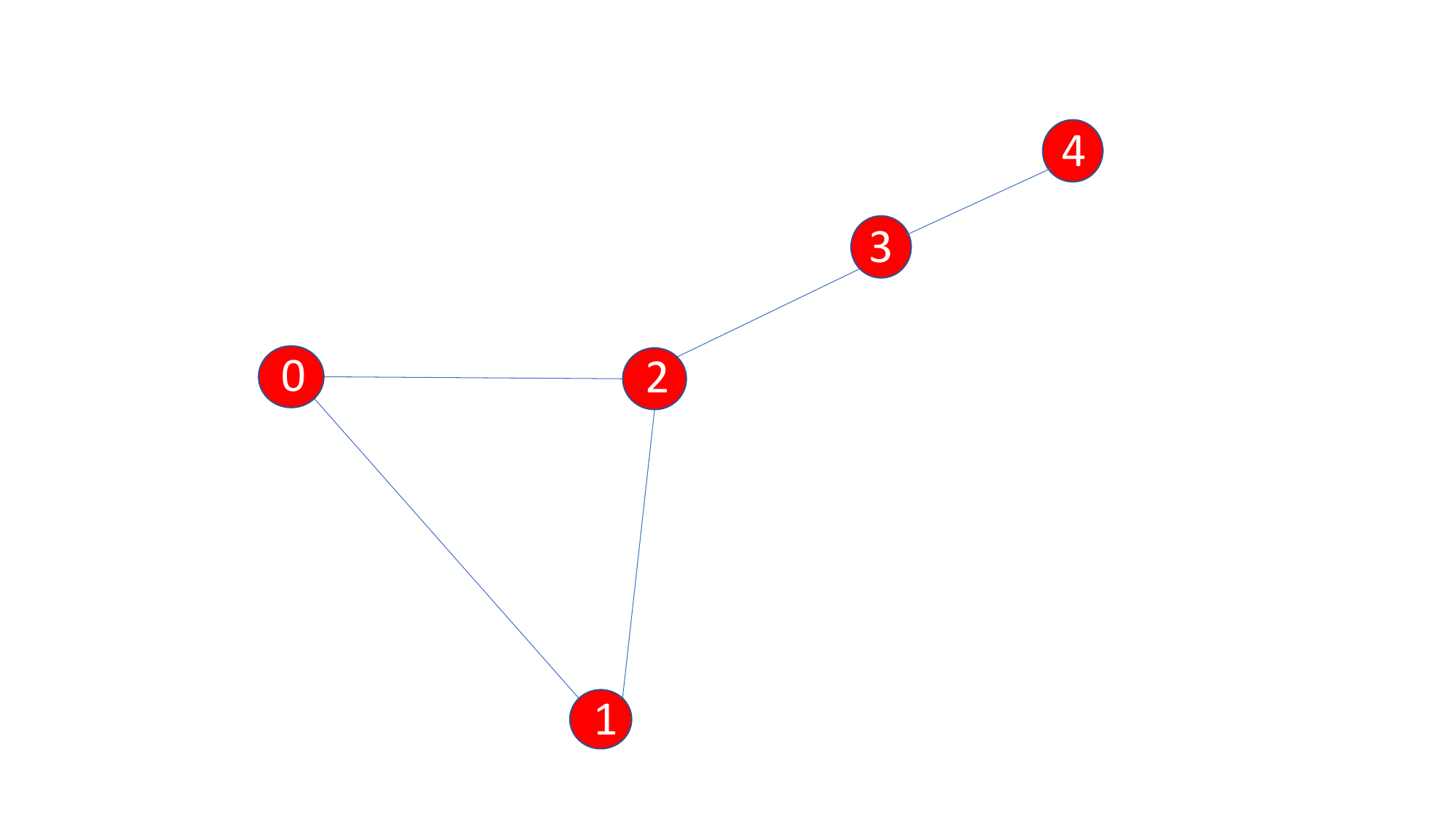}}
\caption{An undirected and unweighted graph. We use node 3 as our target. }
\label{small_graph}
\end{figure}

In this section, we will show how to calculate the hitting time distribution on a small graph using analytical methods. The graph we are going to study contains five nodes, as shown in Fig. \ref{small_graph}.
We take node 3 as our target, and want to calculate the probability of hitting the target starting from each vertex for the first time after exactly $n$ steps. We first illustrate how to calculate the hitting time probability distribution for a simple random walk. For this case, the probability of hitting node 3 starting from node 4, which is an adherent to target node 3, after exactly $n$ steps is 
\begin{align}
P(N_3^{(4)} = n) = \delta_{n, 1}
\end{align}
For nodes 0, 1, 2, we would exploit the Markovian property of the random walk process to calculate the corresponding hitting probabilities. To simplify notation, we encapsulate these hitting probabilities into a column vector as 
\begin{align}
\boldsymbol{X}_{n} = \begin{pmatrix}
P(N_3^{(0)} = n) \\
P(N_3^{(1)} = n) \\
P(N_3^{(2)} = n)
\end{pmatrix}
\end{align}
The probability transition matrix is 
\begin{align}
B = \begin{pmatrix}
0 & 1/2 & 1/2 \\
1/2 & 0 & 1/2 \\
1/3 & 1/3 & 0
\end{pmatrix}
\end{align}
The probability vector satisfies this recursive equation 
\begin{align}
& \boldsymbol{X}_{n} = B \boldsymbol{X}_{n - 1}, n \ge 2, 
\end{align}
with initial condition $\boldsymbol{X}_1 = 
\begin{pmatrix} 
0 & 0 & 1/3
\end{pmatrix}^{\text{T}}$. 
The probability generating function for $\boldsymbol{X}_{n}$ is 
\begin{align}
\label{close_form}
\boldsymbol{\tilde{f}}(z) & = z( I - zB)^{-1} \boldsymbol{X}_1 \\\nonumber
& = \frac{z}{3} \frac{1}{1 - \frac{7}{12}z^2 - \frac{z^3}{6}} \begin{pmatrix}
\frac{z}{2} + \frac{z^2}{4} \\ 
\frac{z}{2} + \frac{z^2}{4} \\ 
1 - \frac{z^2}{4}
\end{pmatrix}
\end{align}
We can easily calculate the first order moments by differentiating the above equation with respect to $z$ at $z = 1$, the results of which are 
\begin{align}
\boldsymbol{\tilde{f}}^{\prime}(1) = \begin{pmatrix}
\langle N_3^{(0)} \rangle \\
\langle N_3^{(1)} \rangle \\
\langle N_3^{(2)} \rangle 
\end{pmatrix} = \begin{pmatrix}
9 \\
9 \\
7 
\end{pmatrix} 
\end{align}

We also implement a Monte Carlo program to calculate the average hitting times. In the Monte Carlo program, we use node 3 as our target node and start each random walk from node 0, 1, and 2. Each random walk terminates at node 3 after some number of steps. By repeating this process thousands of times, we obtain the average number of steps required before the random walker hits the target for the first time. For each node in the set $\{0, 1, 2\}$, we perform $10^6$ random walks. We compare Monte Carlo simulation results with analytical results in Table \ref{small_table}, from which, we can see that the Monte Carlo simulation results are consistent with our analytical results, with relative errors being approximately $10^{-3}$. We do not expect Monte Carlo simulation to give us highly precise numerical results, and the final results of Monte Carlo simulations may vary slightly for different random number generators. 

We also compute the average hitting time starting from each node 0, 1, 2 using numerical methods. We can use either Eq. (\ref{generating_function}) or Eq. (\ref{first_order_moment_power_series}) for this purpose, because the probability transition matrix is small enough for the direct inversion of the matrix to be feasible. However, Eq. (\ref{generating_function}) is no longer practical when the graph is large, and thus even for this small graph, we still prefer to use Eq. (\ref{first_order_moment_power_series}). We implement a Python program on macOS Mojave, and obtain numerical results that are shown together with analytical results and Mont Carlo simulation results in Table \ref{small_table}. It is clear that the numerical results obtained using our algorithm have much higher precision than that of the Monte Carlo simulation results. 

\begin{table}
\begin{tabular}{|c|c|c|c|}
\hline
$\langle N_3^{(i)} \rangle$ & Analytical & Monte Carlo & Numerical \\
\hline
$\langle N_3^{(0)} \rangle$ & 9    	 & 9.00323 			 & 8.99999999999999 \\
$\langle N_3^{(1)} \rangle$ & 9   		 & 8.96693 			 & 8.99999999999999 \\
$\langle N_3^{(2)} \rangle$ & 7   		 & 7.00013  			 & 7.000000000000002 \\
\hline
\end{tabular}
\caption{Simple random walk results for the graph shown in Fig. \ref{small_graph}, using three methods. $\langle N_t^{(i)} \rangle$ is the expected hitting time for a random walk that starts from node $i$ and ends at target node $t$. Here, $t = 3$. }
\label{small_table}
\end{table}

We can obtain similar results for frustrated random walks. For this case, the hitting time probability distribution of node 4 is 
\begin{align}
P(N_{3}^{(4)} = n) = \frac{1}{2^n}
\end{align}
To calculate the hitting time probability distributions starting from nodes 0, 1, 2, we first need to create the probability transition matrix, which is 
\begin{align}
B = \begin{pmatrix}
7/12 & 1/4 & 1/6 \\
1/4 & 7/12 & 1/6 \\
1/6 & 1/6 & 1/2
\end{pmatrix}
\end{align}
And the initial probability vector is $\boldsymbol{X}_{1} = \begin{pmatrix} 0 & 0 & 1/6 \end{pmatrix}^T$. With $B$ and $\boldsymbol{X}_{1}$, we can directly calculate the expected hitting times starting from nodes 0, 1, 2 with node 3 as target, using Eq. (\ref{generating_function}). We can again simulate the frustrated random walk using Monte Carlo method, and see that the exact solution coincides with the Monte Carlo simulation result within tolerance of error. We thus have established the validity of our algorithm for both simple and frustrated random walks, at least for the small graph we are studying.

\subsection{Numerical computation of hitting time distribution on large graphs}
\label{numerical_solution}
\begin{figure}[h!]
\centerline{\includegraphics[width = 0.8 \linewidth]{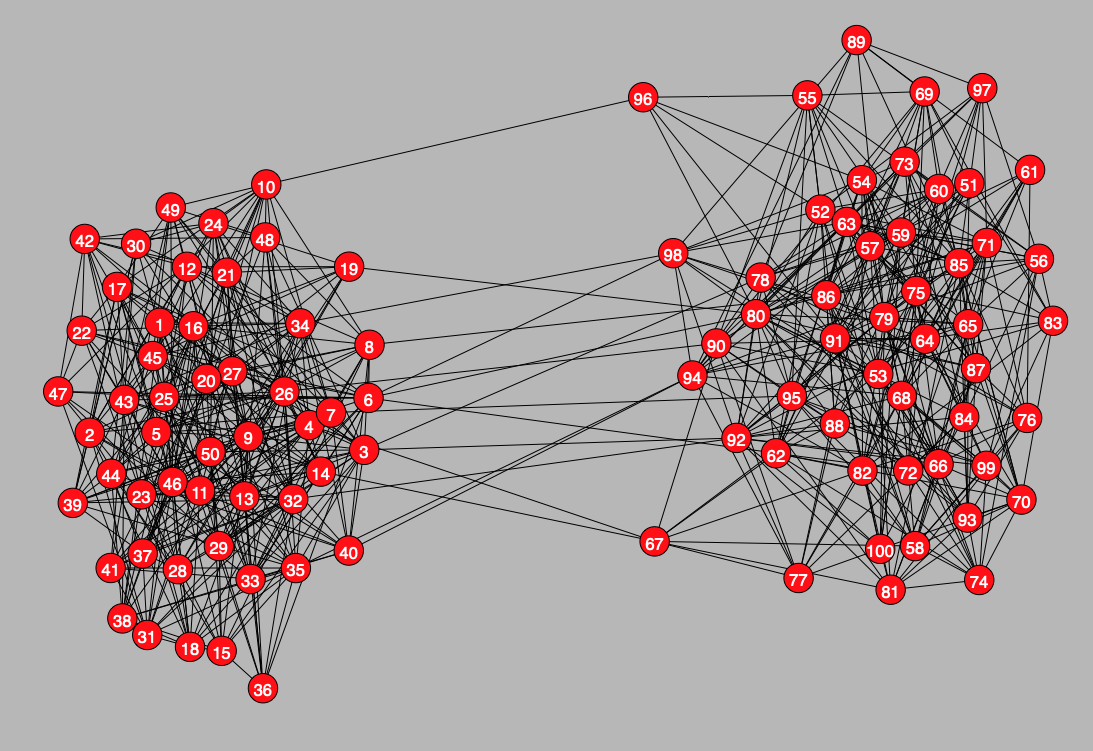}}
\caption{An undirected and unweighted graph with 100 vertices and 704 edges. }
\label{large_graph}
\end{figure}

When dealing with large graphs, it is both tedious and impractical to get an analytical formula like Eq. (\ref{close_form}). Instead, we will use Eq. (\ref{first_order_moment_power_series}) to numerically compute the expected hitting times. Another method to find the hitting time expectations is to use Monte Carlo simulation, although we will see that for a large graph, the running time of Monte Carlo simulation is much longer than the numerical method, thus making the Monte Carlo simulation an inferior alternative. In this section, we apply the numerical method and Monte Carlo simulation method to a connected graph as shown in Fig. \ref{large_graph}. 

\begin{figure}[h!]
\centerline{\includegraphics[width = 0.8 \linewidth]{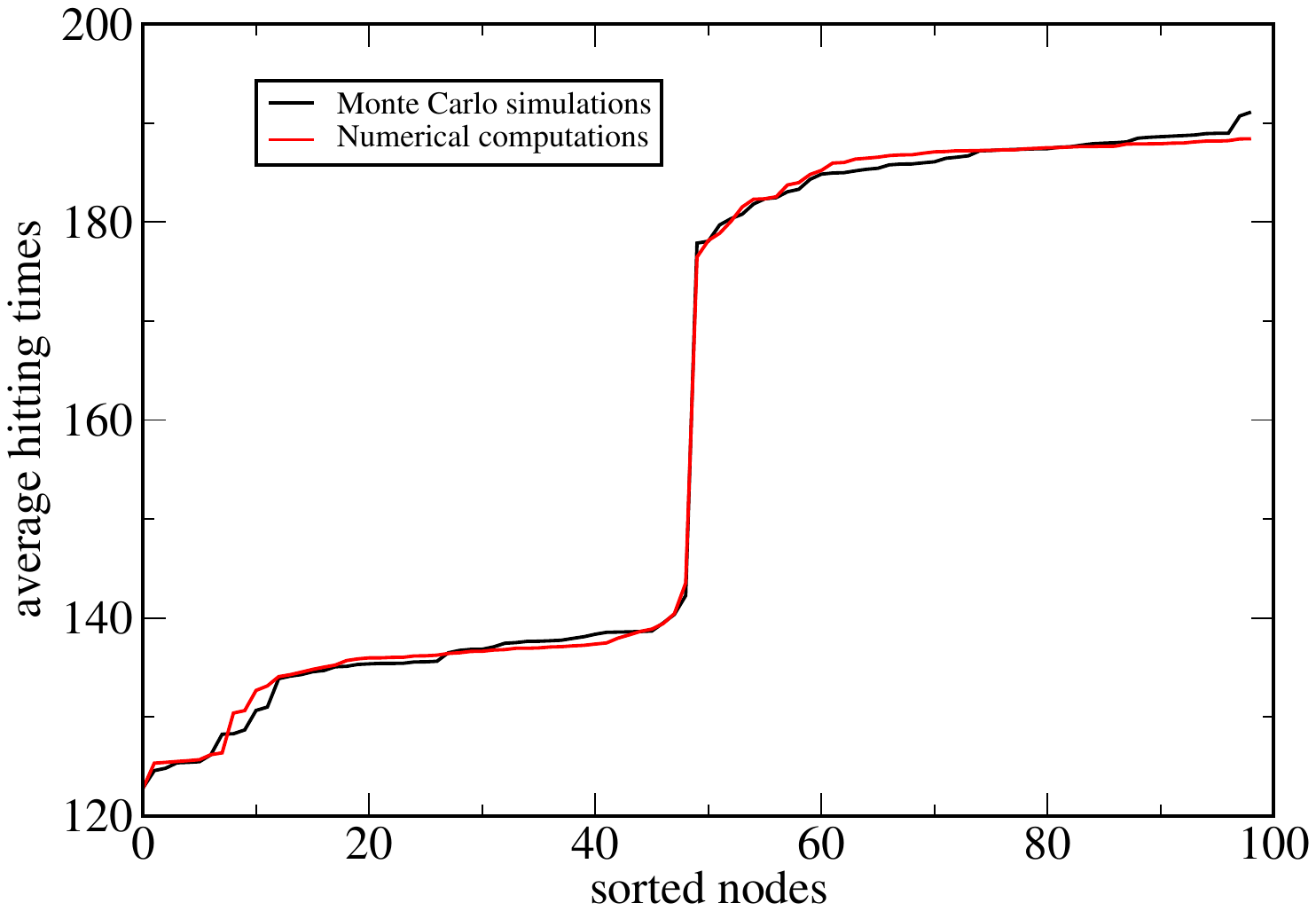}}
\caption{Average hitting time distribution curves of the graph shown in Fig. \ref{large_graph} with node ``1'' as the target.}
\label{sorted_distances}
\end{figure}

We choose node 1 as our target node and calculate the average hitting times starting from all the other nodes in the graph. In order to visualize the results, we sort all the vertices according to their average hitting times relative to the target. In Fig. \ref{sorted_distances}, we plot the results from the Monte Carlo simulation method and the numerical method. We can see that these two methods give almost the same results, although we know that results from Monte Carlo simulations have much lower precision than that obtained from the numerical method. Another weakness of Monte Carlo simulation is that it is much too time-consuming. In order to get the results shown in Fig. \ref{sorted_distances}, we need to run $10^{5}$ random walks from each non-target vertex in the graph, and the whole process costs 163 seconds, whereas in the numerical method, we only need to compute the power series in Eq. (\ref{first_order_moment_power_series}) up to order 7500, and it takes only 3.9 seconds to yield results with machine precision. The larger the graph, the more time-saving the numerical method is compared to Monte Carlo simulations. 

Another feature that is worth mentioning in Fig. \ref{sorted_distances} is that we can distinguish the two communities in the original graph by looking at the distribution of hitting time expectations of the vertices. It is clear that there is a transition region which connects two plateaus in the hitting time expectation vs. sorted vertices curve. The two plateaus correspond to the two communities in the graph. We can interpret the emergence of these two plateaus by imagining that a walker that starts from within one of these two communities tends to get trapped in that community. Once the random walker gets trapped in a community, the hitting time expectations for vertices in that community would not change substantially from vertex to vertex, which gives rise to a plateau in the curve. However, as soon as the random walker finds a bridge leading from one community to another, it will make a rapid transition across the two communities, which results in a significant change in the hitting time expectations. Thus, the calculation of average hitting times gives us a tool for community detection. However, we should make a caveat that this method of community detection is usable only when the number of communities in the graph is small enough and the communities are clearly separated from each other. Or else, this method of community detection is not as good as the ones compiled in Ref. \cite{fortunato2010community, newman2006modularity, PhysRevE.70.066111, PhysRevE.74.036104}.

\end{document}